\documentstyle[aps,prl,floats,psfig]{revtex}

\begin{document}

\draft

\wideabs{

\title{Hole-doped, High-Temperature Superconductors \( Li_{x}BC \), \( Na_{x}BC \) and \( C_{x} \) : A Coherent-Potential-Based Prediction}

\author{Prabhakar P. Singh}

\address{Department of Physics, Indian Institute of Technology, Powai, Mumbai-400076, India}

\date{\today}

\maketitle

\begin{abstract}

Using density-functional-based methods, we show that the hole-doped \( Li_{x}BC \) and \( Na_{x}BC \) in \( P6_{3}/mmc \) crystal structure and \( C_{x} \) in graphite structure are capable of showing superconductivity, possibly with a \( T_{c} \) much higher than that of \( MgB_{2}. \) We use full-potential methods to obtain the optimized lattice constants \( a \) and \( c \), coherent-potential approximation to describe the effects of disorder, Gaspari-Gyorffy formalism to obtain the electron-phonon coupling constant \( \lambda  \), and Allen-Dynes equation to calculate \( T_{c} \) as a function of hole concentration in these alloys.

\end{abstract}

\pacs{PACS numbers: 74.25.Jb, 74.70.Ad} 

} 

The search for new \( MgB_{2} \)-\emph{like} superconductors \cite{nag} has
proved to be elusive so far, in spite of intense experimental effort. The difficulty
in finding new materials with superconducting properties similar to that of
\( MgB_{2} \) \cite{nag,expt,rev} may partly be due to the delicate nature
of interaction responsible for superconductivity (SC) \cite{th,kon,pps01,liu,cho}.
In \( MgB_{2}, \) the relatively high superconducting transition temperature,
\( T_{c} \), of \( 39\, K \) is achieved through a coming together of various
aspects of phonon-mediated electron-electron interaction, such as (i) strong
in-plane \( B-B \) covalent bonding \cite{th,kon,pps01,liu}, (ii) electron
and hole-like cylindrical Fermi sheets along \( \Gamma  \) to \( A \) \cite{th,kon,cho},
(iii) a strong coupling of the \( \sigma  \) states to the in-plane \( B-B \)
bond-stretching mode \cite{th,kon,cho} including an-harmonic effects \cite{liu,cho},
(iv) \( B\, p_{x(y)} \) contribution to the total density of states (DOS) at
the Fermi energy, \( E_{F}, \) and the total DOS at \( E_{F} \) \cite{th,pps01},
and others. 

Most of the experimental as well as theoretical efforts have used \emph{layered
and metallic} materials as the starting point to obtain materials with above-mentioned
electronic properties leading to high-temperature SC but without success. It
may be that we have to follow a different route to obtain new \( MgB_{2} \)-like
superconductors with high \( T_{c}. \) Such an alternative route is provided
by hole-doping of the \emph{layered but semiconducting or insulating} materials
\cite{ros}. 

With the aim of developing materials with \( MgB_{2} \)-like electronic properties,
we have studied how hole-doping can be used to tailor the electronic structure
of \( Li_{x}BC \), \( Na_{x}BC \) and \( C_{x} \) alloys. In particular,
it turns out that the hole-doping makes \( Li_{x}BC \), \( Na_{x}BC \) and
\( C_{x} \) alloys similar to \( MgB_{2} \) in more ways than one, making
these alloys promising candidates for showing high-temperature SC.

In the following, we present the results of (i) fully-relaxed, full-potential
electronic structure calculations for \( LiBC \) and \( NaBC \) in \( P6_{3}/mmc \)
crystal structure and \( C \) in graphite structure, using density-functional-based
methods, (ii) charge-self-consistent, coherent-potential-based electronic structure
calculations for the hole-doped \( Li_{x}BC \), \( Na_{x}BC \) and \( C_{x} \)
alloys, and (iii) the calculations of \( T_{c} \) in hole-doped \( Li_{x}BC \),
\( Na_{x}BC \) and \( C_{x} \) alloys with \( x \) ranging from \( 1 \)
to \( 0.2 \) (hole concentration is equal to \( (1-x) \) ). 

For full-potential calculations, we have used ABINIT code \cite{abinit}, based
on psuedopotentials and planewaves, and optimized the lattice constants \( a \)
and \( c \) of \( LiBC \), \( NaBC \) and \( C \). The effects of hole doping
in \( Li_{x}BC \), \( Na_{x}BC \) and \( C_{x} \) alloys have been studied
using Korringa-Kohn-Rostoker coherent-potential approximation in the atomic-sphere
approximation (KKR-ASA CPA) method \cite{pps94,fau}. For calculating \( T_{c} \)
as a function of \( x \) in \( Li_{x}BC \), \( Na_{x}BC \) and \( C_{x} \)
alloys, we have used the Gaspari-Gyorffy formalism \cite{gas} to evaluate the
electron-phonon (EP) coupling constant \( \lambda  \), and Allen-Dynes equation
\cite{all} to calculate \( T_{c} \). 

Based on our calculations, described below, we find that the hole-doped \( Li_{x}BC \)
and \( Na_{x}BC \) and \( C_{x} \) are capable of showing SC, possibly with
a \( T_{c} \) as high as \( \sim 105\, K \) for \( Li_{0.5}BC \) \cite{ros},
\( \sim 106\, K \) for \( Na_{0.6}BC \) and \( \sim 60\, K \) for \( C_{0.4}. \)
Such a high \( T_{c} \) in these alloys becomes possible partly because the
hole-doping modifies the electronic structure, discussed in terms of the density
of states and spectral function along \( \Gamma  \) to \( A \) direction, and
brings it closer to that of \( MgB_{2} \). 

Before describing our results in detail, we provide some of the computational
details. The structural relaxation of \( LiBC \), \( NaBC \) and \( C \)
was carried out by the molecular dynamics program ABINIT as described in Ref.
\cite{pps01}. The charge self-consistent electronic structure of \( Li_{x}BC \),
\( Na_{x}BC \) and \( C_{x} \) alloys as a function of \( x \) has been calculated
using the KKR-ASA CPA method. We have used the CPA rather than a rigid-band
model because CPA has been found to reliably describe the effects of disorder
in metallic alloys \cite{pps94,fau}, including \( MgB_{2} \) alloys \cite{pps02}.
We parametrized the exchange-correlation potential as suggested by Perdew-Wang
\cite{perdew} within the generalized gradient approximation. The Brillouin
zone (BZ) integrations for charge self-consistency and the DOS calculations
were carried out using \( 1215 \) and \( 4900 \) \textbf{k}-points in the
irreducible part of the BZ, respectively. For DOS (spectral function) calculation,
we added a small imaginary component of \( 1\, (2) \) \( mRy \) to the energy.
The lattice constants for \( Li_{x}BC \), \( Na_{x}BC \) and \( C_{x} \)
alloys as a function of \( x \) were kept fixed to the values given in Table
I. To reduce ASA-related errors in open structures we introduced empty spheres.
The Wigner- Seitz radii for \( Li \) and \( Na \) were slightly larger than
that of \( B \) and \( C \). The maximum \( l \) value used was \( l_{max} \)
= \( 2 \).

\begin{figure}
{\par\centering 
\psfig{file=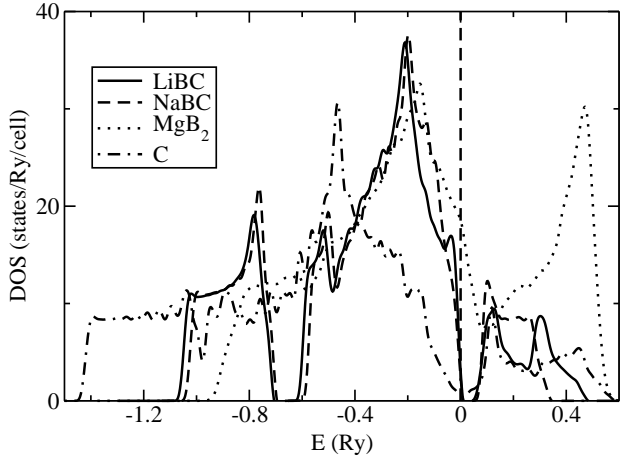,height=7.4cm,angle=-90} 
\par}

\caption{The total density of states calculated at the optimized lattice constants using
the ABINIT program. For comparison the DOS for \protect\( MgB_{2}\protect \)
has been doubled. The vertical dashed line indicates \protect\( E_{F}.\protect \)}
\end{figure}

The \( \lambda  \) was calculated using Gaspari-Gyorffy formalism with the
potentials of \( Li_{x}BC \), \( Na_{x}BC \) and \( C_{x} \) obtained with
the KKR-ASA CPA method. Subsequently, the variation of \( T_{c} \) with \( x \)
was calculated using Allen-Dynes equation. The average value of phonon frequency
\( \omega _{ln} \) for \( MgB_{2} \) was taken from Ref. \cite{kon}, while
the value of \( \mu ^{*} \) was kept fixed at \( 0.09 \). 

\begin{table}

\caption{The calculated lattice constants \protect\( a\protect \) and \protect\( c\protect \)
in atomic units. The experimental lattice constants are shown in the parentheses.}
\begin{tabular}{|c|c|c|}
\hline 
Compound&
 \( a \)&
 \( c \)\\
\hline 
\( LiBC \)&
 5.16 (5.202)&
13.24 (13.342)\\
\hline 
\( NaBC \)&
 5.25&
 15.62\\
\hline 
\( C \) &
 4.63 (4.650)&
 12.62 (12.649)\\
\hline 
\( MgB_{2} \)&
5.76 (5.834)&
6.59 (6.657)\\
\hline 
\end{tabular}\end{table}

\begin{figure}
{\par\centering 
\psfig{file=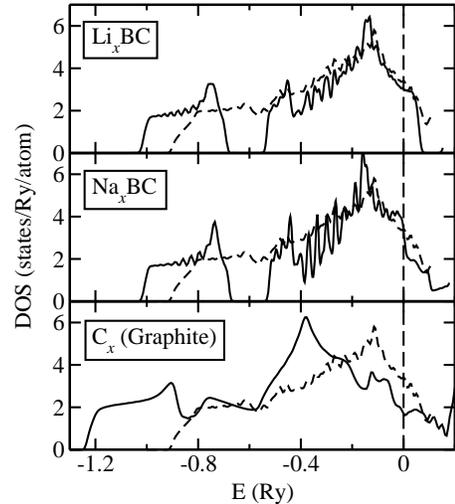,height=7.4cm,angle=-90} 
\par}

\caption{The total DOS (solid line) for \protect\( Li_{x}BC\protect \) (upper panel),
\protect\( Na_{x}BC\protect \) (middle panel), and \protect\( C\protect \)
(lower panel) at \protect\( x=0.5\protect \), calculated using the KKR-ASA
CPA method. For comparison the total DOS for \protect\( MgB_{2}\protect \)
(dashed line) is also shown. The vertical dashed line indicates \protect\( E_{F}\protect \).}
\end{figure}

First, we discuss the changes brought about by hole-doping in the various aspects
of the electronic structure of \( Li_{x}BC \), \( Na_{x}BC \) and \( C_{x} \)
alloys in terms of (i) total DOS, (ii) partial \( B \) and \( C\, p \)-contributions
to the total DOS at \( E_{F} \), and (iii) spectral function along \( \Gamma  \) to
\( A \) evaluated at \( E_{F} \). Then we describe the calculated \( T_{c} \)
for these alloys as a function of hole concentration, which is the main result
of the present work.

We show in Fig. 1 the total DOS for \( LiBC \), \( NaBC \) and \( C \) alloys
calculated using the ABINIT program at the optimized lattice constants as given
in Table I. For comparison we also show the DOS for \( MgB_{2} \) obtained
as described in Ref. \cite{pps01}. The compound \( NaBC \) has a higher \( c/a \)
than \( LiBC \) primarily because of the larger \( Na \) atoms. The difference
in \( c/a \) ratio leads to significant changes between the electronic structure
of \( LiBC \) and \( NaBC \). By comparing the DOS of \( LiBC \), \( NaBC \)
and \( C \) with \( MgB_{2} \), it is clear that if \( E_{F} \) can be moved
suitably inside the valence band then the DOS at \( E_{F} \) in these alloys
can be made comparable to the DOS at \( E_{F} \) in \( MgB_{2}. \) One of
the ways of moving the \( E_{F} \) inside the valence band is through hole-doping,
as described next. 

The changes in the electronic structure of \( Li_{x}BC \), \( Na_{x}BC \)
and \( C_{x} \) alloys as a function of \( x \) has been studied for \( x \)
ranging from \( 1 \) to \( 0.2. \) We find that with increasing hole concentration
\( E_{F} \) moves inside, as expected, and consequently these alloys become
metallic. As an example, we show in Fig. 2 the total DOS of \( Li_{0.5}BC \),
\( Na_{0.5}BC \) and \( C_{0.5} \) alloys, calculated as described earlier.
From a comparison with the DOS of \( MgB_{2} \), calculated using the same
approach and also shown in Fig. 2, it is clear that the DOS around \( E_{F} \)
in \( Li_{0.5}BC \), \( Na_{0.5}BC \) and \( C_{0.5} \) alloys are comparable
to that of \( MgB_{2}. \) Our calculation shows that the hole concentrations
of \( 50\%-80\% \) are ideal for matching the total DOS of \( Li_{x}BC \),
\( Na_{x}BC \) and \( C_{x} \) alloys with that of \( MgB_{2}. \) Knowing
the importance of the total DOS as well as the in-plane \( B\, p \)-contribution
to the total DOS at \( E_{F} \) in \( MgB_{2} \), the alloys \( Li_{x}BC \),
\( Na_{x}BC \) and \( C_{x} \) will be good candidates for showing SC only
if the \( B \) and \( C\, p \)-contributions to the total DOS at \( E_{F} \)
are comparable to that of \( MgB_{2}. \) 

\begin{figure}
{\par\centering 
\psfig{file=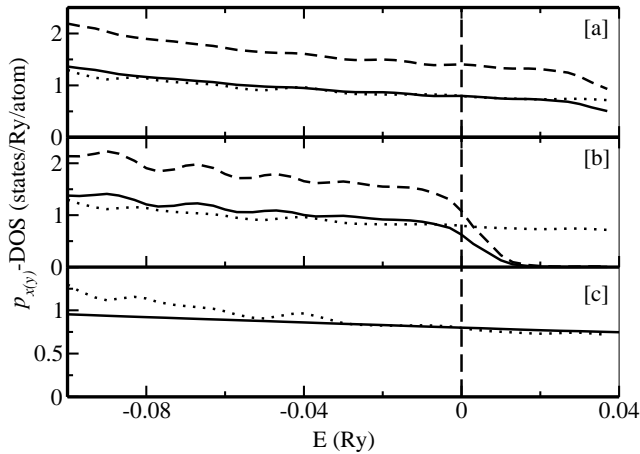,height=7.4cm,angle=-90} 
\par}

\caption{The \protect\( p_{x(y)}\protect \) component of \protect\( B\protect \) (solid
line) and \protect\( C\protect \) (dashed line) partial DOS around \protect\( E_{F}\protect \)
for (a) \protect\( Li_{0.5}BC\protect \), (b) \protect\( Na_{0.5}BC\protect \),
and (c) \protect\( C_{0.5}\protect \), calculated using the KKR-ASA CPA method.
For comparison the \protect\( p_{x(y)}\protect \) component of \protect\( B\protect \)
partial DOS in \protect\( MgB_{2}\protect \) (dotted line) is also shown. The
vertical dashed line indicates \protect\( E_{F}.\protect \)}
\end{figure}

In Fig. 3 we compare the \( p_{x(y)} \) contributions of \( B \) and \( C \)
to the total DOS around \( E_{F} \) with the corresponding contribution from
\( MgB_{2}. \) The \( B\, p_{x(y)} \) contribution in \( Li_{0.5}BC \) is
almost identical to that of \( MgB_{2} \), while the \( C\, p_{x(y)} \) contribution
is slightly higher, indicating a strong \( B-C \) covalent bonding. In \( Na_{0.5}BC \)
the \( B\, p_{x(y)} \) contribution is slightly lower than that in \( MgB_{2}. \)
Thus a further increase in hole concentration will enhance the \( p_{x(y)} \)
contribution to the total DOS. On the other hand in the case of \( C_{0.5} \)
the \( C\, p_{x(y)} \) DOS is identical to that of the \( B\, p_{x(y)} \)
contribution in \( MgB_{2}. \) 

\begin{figure}
{\par\centering 
\psfig{file=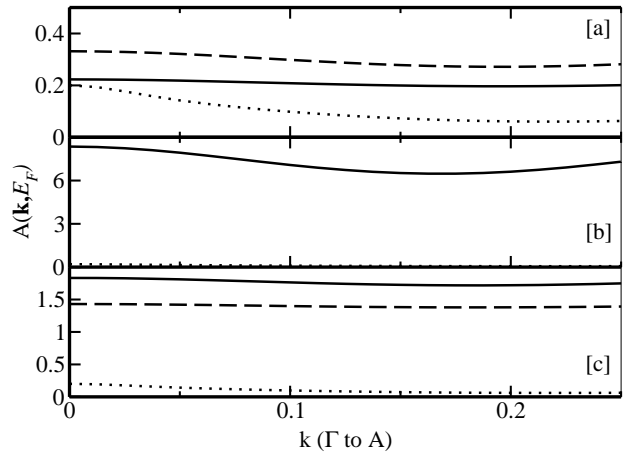,height=7.4cm,angle=-90} 
\par}

\caption{The calculated spectral function along \protect\( \Gamma \protect \) to \protect\( A\protect \)
direction , evaluated at \protect\( E_{F}\protect \), for  \protect\( x=0.3\protect \)
(solid line) and \protect\( x=0.5\protect \) (dashed line) in (a) \protect\( Li_{x}BC\protect \),
(b) \protect\( Na_{x}BC\protect \), and (c) \protect\( C_{x}\protect \) alloys.
For comparison, the spectral function for \protect\( MgB_{2}\protect \) (dotted
line) is also shown. In (b) the spectral function for \protect\( Na_{0.5}BC\protect \)
is out of range of the plot.}
\end{figure}

For \( Li_{x}BC \), \( Na_{x}BC \) and \( C_{x} \) alloys to remain in the
contention for showing \( MgB_{2} \)-like superconducting properties, a favorable
matching of the total DOS and the partial \( B\, p_{x(y)} \) and/or \( C\, p_{x(y)} \)
contributions to the total DOS at \( E_{F} \) must be followed by similarities
in the Fermi surfaces of these hole-doped alloys with that of \( MgB_{2} \),
especially along \( \Gamma  \) to \( A \). A measure of this similarity is
provided by comparing the spectral function, \( A(\mathbf{k},E_{F}) \), of
\( MgB_{2} \) with \( Li_{x}BC \), \( Na_{x}BC \) and \( C_{x} \) alloys
along \( \Gamma  \) to \( A \) \cite{pps02}. Such a comparison is shown in
Fig. 4, where we have plotted \( A(\mathbf{k},E_{F}) \) along \( \Gamma  \) to
\( A \) for \( Li_{x}BC \), \( Na_{x}BC \) and \( C_{x} \) alloys with \( x=0.3 \)
and \( 0.5 \), as well as for \( MgB_{2}. \) We find that \( A(\mathbf{k},E_{F}) \)
for \( Li_{0.5}BC \) is closest to that of \( MgB_{2}, \) indicating similarities
in the Fermi surfaces of the two compounds along \( \Gamma  \) to \( A \).
In \( C_{x} \) the spectral functions for \( x=0.3 \) and \( 0.5 \) are \( 7-8 \)
times larger than the corresponding values in \( MgB_{2} \), while in \( Na_{0.3}BC \)
it is \( 40-45 \) times larger. Based on the comparison of the various aspects
of the electronic structure of \( Li_{x}BC \), \( Na_{x}BC \) and \( C_{x} \)
alloys with \( MgB_{2} \), we find \( Li_{x}BC \) to be the most promising
candidate for showing \( MgB_{2} \)-\emph{like} SC followed by \( C_{x}. \)
Such a conclusion is borne out by our explicit calculation of \( T_{c} \) as
a function of hole concentration in \( Li_{x}BC \), \( Na_{x}BC \) and \( C_{x} \)
alloys.

\begin{figure}
{\par\centering 
\psfig{file=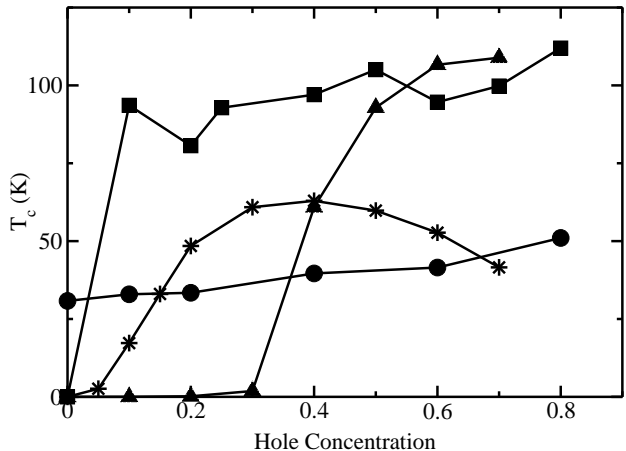,height=7.4cm,angle=-90} 
\par}

\caption{The calculated variation in \protect\( T_{c}\protect \) as a function of hole
concentration for \protect\( Li_{x}BC\protect \) (filled square), \protect\( Na_{x}BC\protect \)
(filled triangle), \protect\( C_{x}\protect \) (star), and \protect\( Mg_{x}B_{2}\protect \)
(filled circle). }
\end{figure}

The calculated \( T_{c} \) as a function of hole concentration in \( Li_{x}BC \),
\( Na_{x}BC \) and \( C_{x} \) alloys is shown in Fig. 5, where we also show
the variation in \( T_{c} \) in \( MgB_{2}. \) Clearly, hole-doped \( Li_{x}BC \),
expected to have a \( T_{c} \) of around \( 105\, K \) with \( 50\% \) hole
concentration, offers the best hope for improving the \( T_{c} \) of \( MgB_{2} \)-type
superconductors. The EP coupling is much stronger in \( Li_{x}BC \) (\( \lambda \sim 1.7 \)
for \( x=0.5 \)) than in \( MgB_{2} \) or any other alloys that we have considered.
Our results for \( Li_{x}BC \) are consistent with the results of Ref.\cite{ros}.
Our calculation also shows that \( Na_{x}BC \), \emph{if it can be synthesized},
will also have a \( T_{c} \) comparable to that of \( Li_{x}BC \) albeit at
a higher hole concentration. The change in \( T_{c} \) of \( MgB_{2} \) with
increasing hole concentration is not large. Surprisingly, \( C_{x} \) is also
found to show SC at elevated temperature.

The work presented here can be further improved by optimizing the lattice constants
\( a \) and possibly \( c \) \cite{vit}, within the CPA, as a function of
hole concentration. However, the results presented in this paper are very robust
and, in our opinion, a slight change in lattice constants due to optimization
will not change the main conclusions of the present work.

In conclusion, we have presented the results of a fully-relaxed, full-potential
electronic structure calculations for \( LiBC \) and \( NaBC \) in \( P6_{3}/mmc \)
crystal structure and \( C \) in graphite structure. Using CPA, we have discussed
the effects of hole-doping in \( Li_{x}BC \), \( Na_{x}BC \) and \( C_{x} \)
alloys in terms of (i) total DOS as well as \( B\, p \) contribution to it
and, (ii) the spectral function along \( \Gamma  \) to \( A \), and shown
that the electronic structure of the hole-doped alloys is similar to that of
\( MgB_{2}. \) Our calculated \( T_{c} \) clearly shows \( Li_{x}BC \) and
\( C_{x} \) to be superconducting in the range \( 0.2\leq x\leq 0.8 \), and
\( Na_{x}BC \) to be superconducting in the range \( 0.2\leq x\leq 0.65 \),
with \( Li_{0.5}BC \) showing a possible \( T_{c} \) of \( 105\, K. \)

\end{document}